\documentclass[preprint,12pt]{elsarticle}

\usepackage{graphics}
\usepackage{amsmath}
\usepackage{fancyhdr}
\usepackage{amssymb}
\usepackage{latexsym}
\usepackage{graphicx}
\usepackage{epsfig}
\usepackage{setspace}
\usepackage{graphicx}
\usepackage{epstopdf}

\newcommand{\beq}{\begin{equation}}
\newcommand{\bseqs}{\begin{subequations}}
\newcommand{\eseqs}{\end{subequations}}
\newcommand{\balign}{\begin{align}}
\newcommand{\ealign}{\end{align}}
\newcommand{\eeq}{\end{equation}}
\newcommand{\beql}{\begin{equation} \label}
\newcommand{\beqs}{\begin{eqnarray}}
\newcommand{\eeqs}{\end{eqnarray}}
\newcommand{\beas}{\begin{eqnarray*}}
\newcommand{\eeas}{\end{eqnarray*}}
\newcommand{\ber}{\begin{array}}
\newcommand{\eer}{\end{array}}
\newcommand{\becs}{\begin{cases}}
\newcommand{\eecs}{\end{cases}}

\newcommand{\leftm}{\left[\begin{array}}
\newcommand{\rightm}{\end{array}\right]}

\newcommand{\dT}{{{\delta T}}}
\newcommand{\bbm}{\begin{bmatrix}}
\newcommand{\ebm}{\end{bmatrix}}

\newcommand{\iif}{{{\rm if}}}

\newcommand{\iin}{{{\rm \;in\;}}}

\newcommand{\carnot}{{\rm carnot}}

\newcommand{\metal}{{\rm mt}}
\newcommand{\semi}{{\rm sm}}
\newcommand{\gain}{{\rm gain}}

\newcommand{\tot}{{\rm tot}}
\newcommand{\out}{{\rm out}}

\newcommand{\ex}{{\rm ex}}
\newcommand{\iinn}{{\rm }}
\newcommand{\pipe}{{\rm pp}}
\newcommand{\dP}{{\vartriangle \!\! P}}

\newcommand{\tube}{{\rm tb}}

\begin{document}

\begin{frontmatter}



\title{Large-scale Ocean-based or Geothermal Power Plants by Thermoelectric Effects}


\author{Liping Liu}

\address{{Department of Mechanical \& Aerospace Engineering, Rutgers University,Piscataway, NJ 08854, USA}\\
{Department of Mathematics, Rutgers University, Piscataway, NJ 08854, USA}}

\begin{abstract}
Heat resources of small temperature difference are easily accessible, free and unlimited on earth. Thermoelectric effects provide the technology for converting these heat resources directly into electricity. We present designs of electricity generators based on thermoelectric effects and using heat resources of small temperature difference, e.g., ocean water at different depths and geothermal sources, and conclude that large-scale power plants based on thermoelectric effects are feasible and economically competitive. The key observation is that the power factor of thermoelectric materials, unlike the figure of merit, can be improved by orders of magnitude upon laminating good conductors and good thermoelectric materials. The predicted large-scale power plants based on thermoelectric effects, if validated, will have a global economic and social impact for its scalability, and the renewability, free and unlimited supply of heat resources of small temperature difference on earth.

\end{abstract}


\end{frontmatter}


\section{Introduction}
\label{}

By the Second Law of thermodynamics, heat, as a form of energy, can be converted into usable energy only if there exists a temperature different between two heat reservoirs. The ideal efficiency of conversion is given by the Carnot efficiency:
\beqs \label{eq:carnot}
\eta_{\carnot}=\frac{T_h-T_c}{T_h},
\eeqs
where $T_h$ and $T_c$ are the high and low temperature, respectively.  From this viewpoint, the largest accessible energy reservoir on earth is undoubtedly the oceans in regard of the thermocline in oceans: the surface water temperature within 100 meters of sea level is around 20K higher than that below 600 meters in tropical regions since surface water is heated by, stores and concentrates solar energy. If this energy resource can be converted to electricity at a competitive cost, even with a fraction of the ideal Carnot efficiency, the energy problem in the world is  solved once and forever.

\begin{figure}[h]
\centering
\includegraphics[scale=0.8]{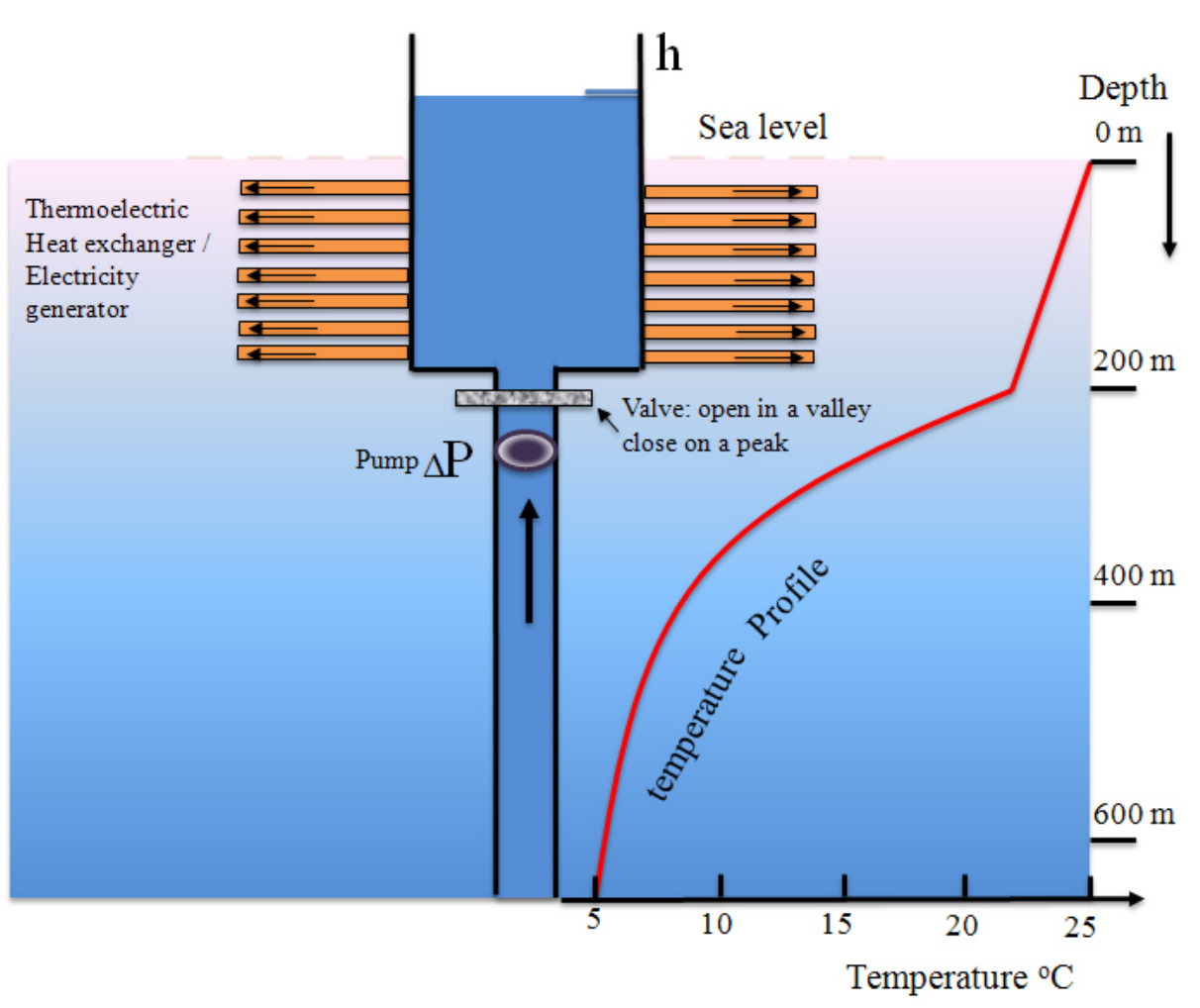}
\caption{\small  A thermoelectric power plant using temperature difference of ocean water at different depths: cold water is pumped and passes through the heat exchanger / generator where it is heated by surface water and generates electricity. Energy of ocean waves can be easily harvested to drive the flows. Surface water has higher temperature since it absorbs, stores and concentrates sunlight energy. The temperature profile is sketched according to~\cite{NOAAthermocline} in tropical regions.}\label{fig:Oceangenerators}
\end{figure}
\renewcommand{\labelenumi}{(\roman{enumi})}
 Conventional technologies of large-scale power plants burning chemical or nuclear fuels have to first convert heat into kinetic energy, which is then used to propel electromagnetic power generators. These power plants cannot make use of  heat  resources of small temperature difference. Thermoelectric (TE) materials, on the other hand, can directly convert heat into electricity no matter how small the temperature difference is \cite{Disalvo1999}.  Small-scale TE generators have been commercialized in applications of microelectronics for reliable and independent power \cite{Leonovetal2007, LeonovVullers2009}, automobiles for improving overall fuel efficiency \cite{Saqretal2008}, and electricity supply in remote areas \cite{Nuwayhidetal2003}. In conventional designs, the efficiency of conversion is a predominant design criterion to control cost since fuels account for the major cost of power generation \cite{Mahan1998, Yangetal2000, Rowe2006,NolasPoonKanatzidis2006, SnyderToberer2008, Lanetal2010}. The efficiency of a TE generator is dictated by a material property, namely, the dimensionless figure of merit which, to our best knowledge, is hard to improve, and the state-of-the-art TE materials have a dimensionless figure of merit of around $2.4$ at the room temperature~\cite{Venkatasubramanianetal2001, Rowe2006}, corresponding to an efficiency of about 20\% of the ideal Carnot efficiency. A larger dimensionless figure of merit is necessary for large-scale TE generators to be competitive against conventional conversion methods if fuels are consumed, which motivates recent renewed interest in thermoelectric materials focusing on improving the conversion efficiency.

\begin{figure}[t]
\centering
\includegraphics[scale=0.9]{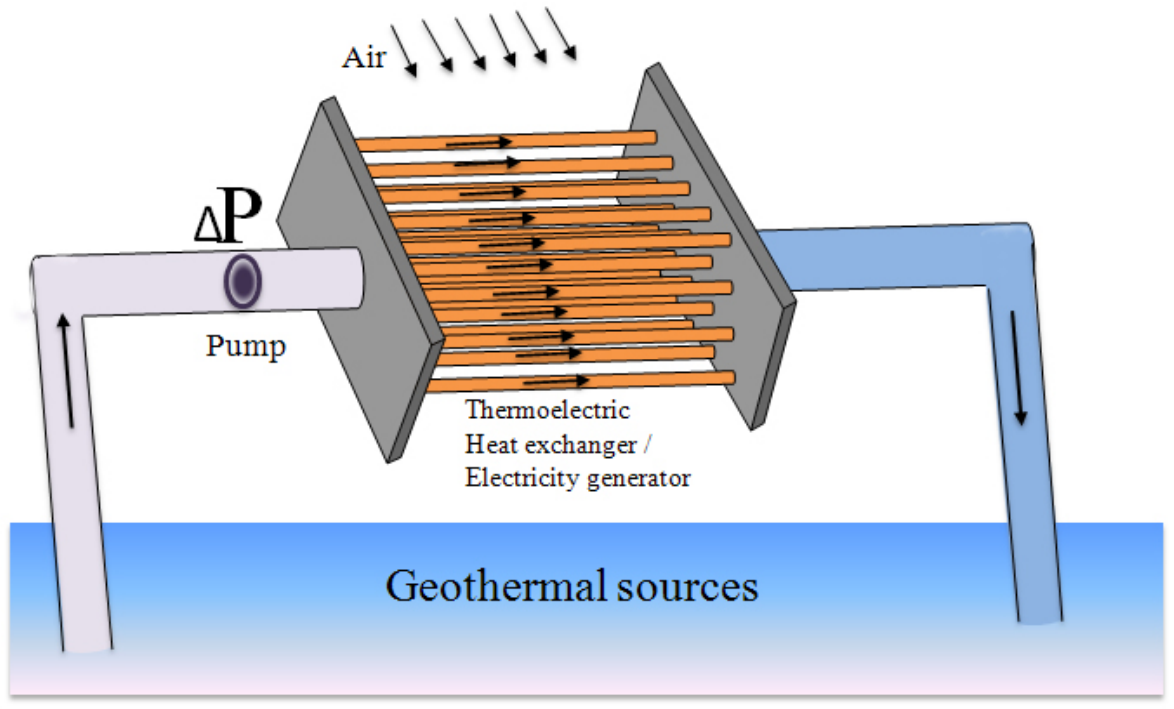}
\caption{\small A thermoelectric generator using geothermal resources. Hot water is pumped and passes through the heat exchanger where it is cooled by air and generates electricity.}\label{fig:geogenerators}
\end{figure}

Here we present  designs of  TE generators at the scale of megawatt. The heat resources are taken as free and unlimited, e.g., ocean water at different depths (Fig. \ref{fig:Oceangenerators}, \cite{NOAAthermocline}),  geothermal sources (Fig. \ref{fig:geogenerators}, \cite{DOEgeothermal}), or waste heat from conventional power plants \cite{Hsuetal2011}. The main cost of the generator comes from the materials, the fabrication and assembly processes of the generator, and possibly the energy needed to drive the flows for heat exchange and electricity generation. This cost basis, as will be shown below, is realistic and shifts the importance of conversion efficiency (figure of merit) to conversion capacity (power factor), and favors simple structures for ease of mass production. Power factors of thermoelectric materials, unlike figure of merit, can be improved by orders of magnitude  by engineering thermoelectric structures, e.g., laminating two or more TE materials.

Below we focus our analysis on large-scale  ocean-based TE power plants using water at different depths as ``fuel'' for its renewability and scalability.  As illustrated in Fig.~\ref{fig:Oceangenerators}, cold  water is pumped and driven through tubes in a heat exchanger where it is  heated by surface water. The tubes are made of TE materials, and therefore, as heat exchanges across the TE tube walls, electricity is generated. In spite of the well-known shortcoming of low conversion efficiency,  large-scale ocean-based TE power plants enjoy the following unique and unprecedented advantages:
\begin{enumerate}
\item Heat resources of small temperature difference,  i.e.,  water at deep sea (below 600m),  are easily accessible,  free and unlimited.
    The power cost to pump water and drive the flows can be made free by harvesting energy from ocean waves.  A valve can be designed to open  in a valley  (below sea level) and close on a peak (above sea level) of  ocean waves, and henceforth an average water height above sea level in the cold water tank can be maintained without any power cost.

\item It costs no land and can be installed in three dimensions, and hence has superior scalability.

\item It  is green, reliable since there is no  moving solid parts, and has low maintenance cost.

\item The power output is independent of the hour of a day and season at a tropical site.
\item It counteracts global warming.
\end{enumerate}

In addition we notice that   ocean waves and convection naturally mix surface water and hence no power is needed to drive the  flows exterior to the heat exchanger, and that there is no stringent  structural requirement on the long underwater pipe if the material density is close to that of water. Structural materials of this kind can be realized by reinforced foam or plastics at a low cost which also serve as good thermal insulators.

\section{Proposed Designs of the generator}
Below we show large-scale ocean-based TE power plants at the order of megawatt are feasible and economically competitive even if  power is needed to pump water and drive the flows on a waveless day. To see this and for simplicity, we assume the flows in the underwater pipe and TE tubes are laminar, and classic fluid dynamics asserts that the power needed to pump water to a height $h$ and the flow rate through TE tubes driven by a pressure $\rho g h$ is given by \cite{FoxPritchardMcDonald2008}:
\beqs \label{eq:Pd}
&&P_{\rm loss}= \mu \,  \gamma_\pipe N_\pipe   L_\pipe \frac{  Q_\pipe^2 }{  A^2_\pipe } + \mu \,  \gamma_\tube N_\tube   L_\tube\frac{  Q_\tube^2 }{  A^2_\tube },\\
&&Q_\tube=\frac{\rho g h A_\tube^2}{\mu \gamma_\tube L_\tube} ,
\eeqs
where  where $\mu=1.0 \times 10^{-3} $ Pa$\cdot$s ($\rho =10^3 kg/m^3$) is the viscosity (density) of water,  $N_\pipe$ ($N_\tube$) is the number of underwater long pipes (near surface short TE tubes),  $L_\pipe$ and $A_\pipe$ ($L_\tube$ , $A_\tube$) are the length and cross-sectional area of the pipe (TE tube), $Q_\pipe$ ($Q_\tube$) is the volumetric flow rate through the pipe (TE tube),  and $\gamma_\pipe$ ($\gamma_\tube$) is a dimensionless geometric factor depending on the cross-sectional shape of  the pipe (TE tube).  If both are circular, $\gamma_\pipe=\gamma_\tube=8\pi$. Further,   $N_\pipe Q_\pipe=N_\tube Q_\tube=Q$ is the total volumetric flow rate through the generator, the second term in $P_{\rm loss}$ can be identified as $\rho g h Q$, and $P_{\rm loss}$ scales as $1/N_\pipe$ and $1/N_\tube$ for fixed  $Q$. The formula~\eqref{eq:Pd} underestimates the actual power loss for neglecting turbulence, friction with the walls, etc, but  is accurate at the order of magnitude and sufficient for our subsequent analysis.

\begin{figure}[h]
\begin{center}
    {\scalebox{0.9}{\includegraphics{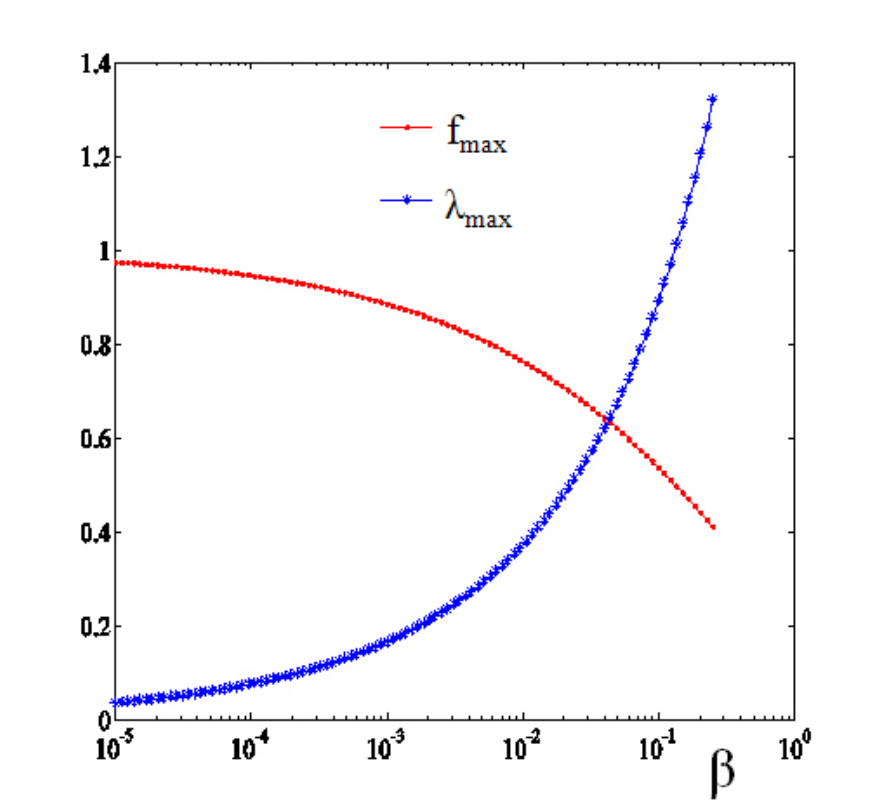}}}

\caption{The maximum dimensionless factor $f_{\max}(\beta)$ and the corresponding maximizer $\lambda_{\max}(\beta)$. }\label{fig:ffactor}

\end{center}
\end{figure}

Electrically, the TE tube shall be divided into segments to boost electric potential and maximize power output as illustrated in Fig.~\ref{fig:connections} (a)-(b). The actually geometry and structure of the TE tube can be engineered to have various cross-sectional shapes, longitudinal connections, etc, which will not be considered here. The interested reader is referred to \cite{SuzukiTanaka2003, Suzuki2004, Snyder2004} Then  one can find that each TE tube has a power output given by (Appendix):
\beas
P_\out=\frac{ P_f \dT^2 L_\tube l_\tube}{4t_\tube  } \cdot \frac {1-\exp(-\lambda)}{\lambda},
\eeas
where $l_\tube$ is the (wetted) perimeter of the TE tube,
$\lambda=\gamma_\tube  \frac{ 2 \mu \kappa_q }{C_p\rho^2 g}\cdot \frac{1}{h} \cdot \frac{l_\tube L_\tube^2 }{t_\tube  A_\tube^2}$, $C_p=4.18\times 10^3\,J/kg\cdot K$ is the specific heat of water, $P_f=\sigma s^2$ is the power factor of the TE material, $\kappa_q=\kappa+T_0\sigma s^2/2$, and $\kappa$, $\sigma, s$ is the thermal conductivity, electric conductivity, Seebeck coefficient of the TE material, respectively. Therefore, in account of the  power loss $P_{\rm loss}$ (cf., \eqref{eq:Pd}), the power gain from each TE tube is given by
\beqs \label{eq:Pgain}
P_\gain=P_\out-  \mu\, \gamma_\tot  L_\tube  \frac{  Q_\tube^2 }{ A^2_\tube }
=\frac{P_f \dT^2 l_\tube L_\tube}{4t_\tube } f(\lambda, \beta),
\eeqs
 where
 \beqs
 f(\lambda, \beta)=\frac{1-\exp(-\lambda)}{\lambda}- \frac{\beta}{\lambda^2},
\eeqs
$
\gamma_\tot=\gamma_\tube+\frac{\gamma_\pipe L_\pipe A_\tube^2 N_\tube }{\gamma_\tube L_\tube A_\pipe^2 N_\pipe},
$
$\beta = \gamma_\tot  \frac{16 \mu \kappa_q^2 }{\rho^2 C_p^2  P_f \dT^2}\cdot \frac{L_\tube^2 l_\tube}{t_\tube  A_\tube^2}$ and hence $\lambda=\frac{1}{h} \cdot \frac{\gamma_\tube}{\gamma_\tot}\cdot\frac{C_pP_f\dT^2}{8 \kappa_q g} \beta.$

\begin{figure}[h]
\begin{center}
    {\scalebox{0.9}{\includegraphics{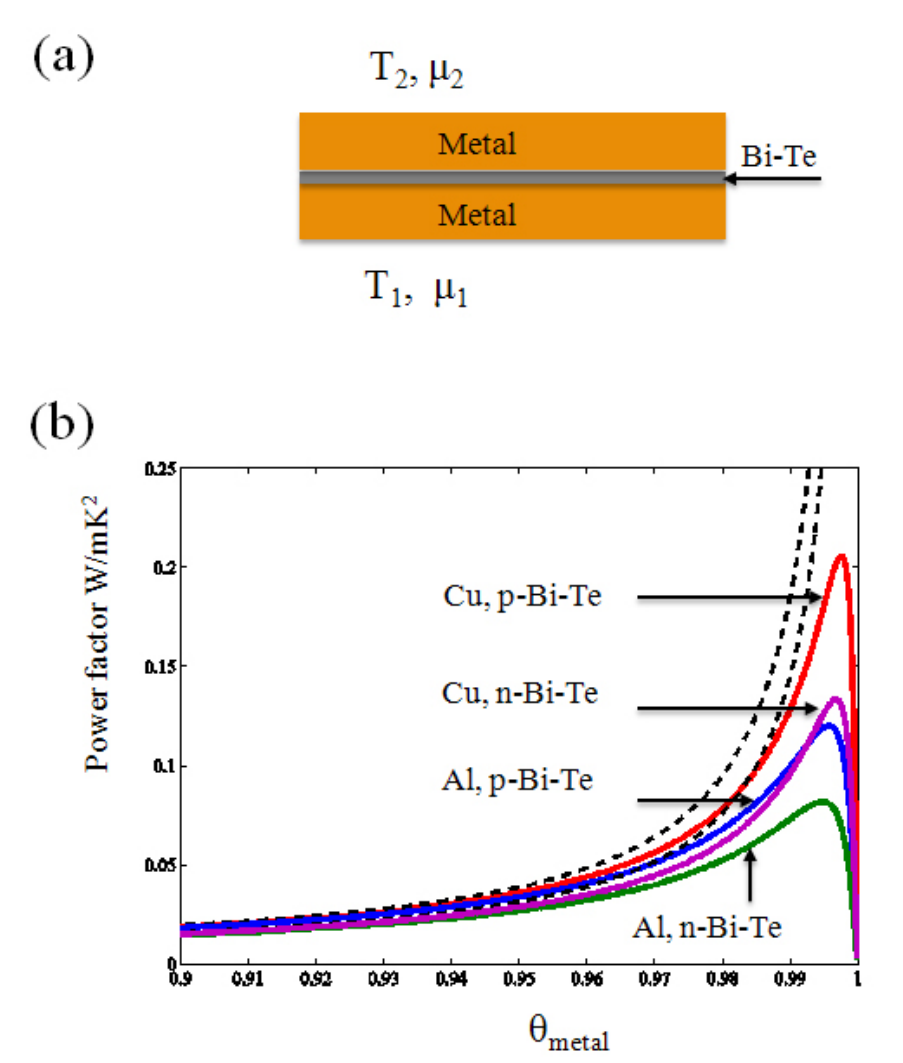}}}

\caption{Power factors of simple laminates of two materials: (a) a sketch of the laminate, and (b) the power factor as a function of volume fraction of the metal. The solid curves are plotted according to the self-consistent continuum theory, i.e., \eqref{eq:Pfe2}; the dashed lines are plotted according to the rough estimate \eqref{eq:Pfe1}. There is no discernible difference between two predictions for lower volume fraction of metal.  }\label{fig:pfdesign}

\end{center}
\end{figure}

From the above equation it is clear that  a large power factor  $P_f$  is crucial to improve $P_\gain$. The power factor can be significantly improved by composites. To see this, we consider  laminates of two materials as illustrated in Fig.~\ref{fig:pfdesign} (a). From the definition $P_f=\sigma s^2$, we see that the power factor will be improved if electric conductivity increases while Seebeck coefficient $s$ is retained.  We therefore consider  laminates of  good conductors, e.g., Cu ($\sigma=5.85\; e7/\Omega m$, $\kappa=401\,W/m K$, $s=1.90 \mu V/K$) or Al ($\sigma=3.66\; e7/\Omega m$, $\kappa=237\,W/m K$, $s=-1.66 \mu V/K$)~\cite{Rowe2006}, with  TE materials with large Seebeck coefficients, e.g.,  Bi-Te semiconductor systems of p-type doped (Bi$_{0.25}$Sb$_{0.75}$)$_2$ Te$_3$  ($\sigma=0.33\; e5/\Omega m$, $\kappa=0.559\,W/m K$, $s=245 \mu V/K$) or n-type doped Bi$_2$( Te$_{0.94}$Sb$_{0.06}$)$_3$ ($\sigma=0.36\; e5/\Omega m$, $\kappa=0.788\,W/m K$, $s=-209 \mu V/K$) \cite{YamashitaOdahara2007}.

Since the electric conductivity and thermal conductivity  between the metals and semiconductors differ by more than two orders of magnitude, for the applied boundary conditions illustrated in Fig.~\ref{fig:pfdesign}~(a) we anticipate that  electric potential and temperature drop mainly occurs across the semiconductor layer, and hence the Seebeck coefficient remains roughly to be that of the semiconductor, i.e., the effective Seebeck coefficient $s^e=s_\semi$. For uncoupled electric or thermal conductivity problems, it is classic that the effective conductivities of laminates are given by
\beas
\sigma^e=(\frac{\theta_\metal}{ \sigma_\metal}+\frac{\theta_\semi}{ \sigma_\semi})^{-1}, \qquad \kappa^e=(\frac{\theta_\metal}{ \kappa_\metal}+\frac{\theta_\semi}{ \kappa_\semi})^{-1},
\eeas
and hence  the effective power factor is given by
\beqs \label{eq:Pfe1}
P_f^e=\sigma^e (s^e)^2=(\frac{\theta_\metal}{ \sigma_\metal}+\frac{\theta_\semi}{ \sigma_\semi})^{-1} (s_\semi)^2,
\eeqs
where $\theta_\metal$ and $\theta_\semi$ are the volume fractions of metal and semiconductor in the laminate, respectively. The above formula predicts that the power factor would increase significantly if $\theta_\metal$  increases since $\sigma^e$ increases monotonically from that of the semiconductor to that of the metal as $\theta_\metal$ increases from 0 to 1.

Equation~\eqref{eq:Pfe1} is accurate when $\theta_\metal$ is not close to one. On the other hand, it is desirable to have $\theta_\metal$ close to one to improve power factor as much as possible. In account of electric and thermal couplings, a  continuum theory for TE materials has been systematically developed in Liu (2012), where  the effective TE properties of the laminate are predicted as
\beqs \label{eq:Pfe2}
&&\begin{bmatrix}
T_0\sigma^e&T_0^2 \sigma^e s^e\\
T_0^2 \sigma^e s^e&T_0^2 [\kappa^e+T_0\sigma^e (s^e)^2]\\
\end{bmatrix}=\\
&& \qquad\bigg\{ \theta_\metal
\begin{bmatrix}
T_0\sigma_\metal &T_0^2 \sigma_\metal s_\metal\\
T_0^2 \sigma_\metal s_\metal&T_0^2 [\kappa_\metal+T_0\sigma_\metal (s_\metal)^2]\\
\end{bmatrix}^{-1} \nonumber \\
&&\qquad +\theta_\semi \begin{bmatrix}
T_0\sigma_\semi&T_0^2 \sigma_\semi s_\semi\\
T_0^2 \sigma_\semi s_\semi&T_0^2 [\kappa_\semi+T_0\sigma_\semi (s_\semi)^2]\\
\end{bmatrix}^{-1}\bigg\}^{-1}. \nonumber
\eeqs
Figure~\eqref{fig:pfdesign} (b) shows the power factors of Cu and p-type doped ( Bi$_{0.25}$Sb$_{0.75}$)$_2$Te$_3$, Cu and n-type doped Bi$_2$( Te$_{0.94}$Sb$_{0.06}$)$_3$, Al and p-type doped (Bi$_{0.25}$Sb$_{0.75}$)$_2$Te$_3$, Al and n-type doped Bi$_2$( Te$_{0.94}$Sb$_{0.06}$)$_3$ as a function of volume fraction $\theta_\metal$. We observe that there exists an optimal volume fraction, typically very close to one, such that the power factor is maximized with the maximum power factor about 100 times larger than the constituent semiconductor. In particular, for  Cu and p-type doped ( Bi$_{0.25}$Sb$_{0.75}$)$_2$Te$_3$, the sandwich laminate has maximum power factor $P^e_f=0.206\,W/mK^2$ ($\sigma^e=8.30 e6 \,/\Omega m$, $\kappa^e=158 \,W/ mK$, $s^e=145\, \mu V/K$) when $\theta_{\rm Cu}=0.9975$. We also remark that the sandwich structure illustrated in Fig.~\ref{fig:pfdesign}(a) is convenient for mass-production and  for electric connections since the metal can serve as electrodes as well.

Further, from \eqref{eq:Pgain} it is clear that small $\beta$ is favored for large $f(\lambda, \beta)$. We shall assume $\beta\le 1$ by tuning geometric factor $L_\tube , l_\tube, A_\tube$. Upon maximizing $f(\lambda, \beta)$ over $\lambda>0$ for given $\beta$, we obtain the maximum $f_{\max}(\beta)$ and maximizer $\lambda_{\max}(\beta)$, which are shown in Fig.~\ref{fig:ffactor} (c).
From Fig.~\ref{fig:ffactor} (c) we observe that $f_{\max}$ depends on $\beta$ weakly, varies from $1$ to $0.1$ when $\beta$ varies from $10^{-5}$ to $1$. Moreover, by~\eqref{eq:Pgain} the power gain from each TE tube $P_\gain$ scales as $1/t_\tube $. Therefore, a thin tube wall is preferred and  the smallest $t_\tube $ is limited by the manufacture and structure requirement which may be reasonably chosen as $t_\tube =0.1\sim 1\, mm$.

To fix the design parameters, we select as the TE tube wall material the sandwich laminate of Cu and p-type doped (Bi$_{0.25}$Sb$_{0.75}$)$_2$Te$_3$ with the maximum power factor $P_f^e$. For maximum power factor, the thickness of semiconductor layer between metals shall be chosen as $t_\semi=t_\tube  * 0.25\%= 0.25 \sim 2.5\, \mu m$, which does not appear to be a challenge for modern technology of nanofabrication. Assume that $\dT=10$K which suffices to account for the heating up of cold water when being pumped up from deep sea. Then the dimensionless parameters $\beta, \lambda$ are given by (SI units):
\beas
\beta=1.38 \times 10^{-12} \times \gamma_\tot \frac{l_pL_\tube^2}{t_\tube  A_\tube^2}, \qquad \lambda=\frac{6.28 \gamma_\tube}{\gamma_\tot} \times\frac{\beta}{h}.
\eeas
 Assume both the long underwater pipe and short TE tubes are circular and select $R_\pipe=1$ m, $L_\pipe=10^3$ m, $N_\pipe=1$, $R_\tube=10^{-2}$ m, $L_\tube=1$ m, $N_\tube\le 10^5$, and hence  $\gamma_\tot\approx \gamma_\tube=8 \pi$, $\beta=2.21\times 10^{-5}/t_\tube $. Upon choosing $h$ such that $\lambda=\lambda_{\max} (\beta)$, we find the power gain $P_\gain=0.324 \cdot f_{\max}/t_\tube $ Watt  from each TE tube and the number of TE tubes needed for a megawatt power gain which are listed in Table~\ref{tab:optmp1}. We remark that rectangular fin-like TE tube may be more favorable in practice for ease of fabrication and mass-production which, however, has little effect on our prediction of power gain from each TE tube.

\begin{center}

{\footnotesize {\bf Designs of a megawatt TE generator ($\dT=10$K)}. \newline
Physical means and dimensions of the symbols: $t_\tube , R_\tube, L_\tube$ --- thickness, inner radius and length of TE tube in mm, cm, m, respectively; $f_{\max}$ --- the maximum dimensionless factor in \eqref{eq:Pgain}; $h$ in meter --- the pressure needed to drive the flow in $\rho g h$ (cf., Fig.~\ref{fig:Oceangenerators}); $P_\gain$ --- the power gain from a tube in KW; $N_\tube$ --- the number of tubes needed for one megawatt power gain.  }

\vspace{-0.5cm}
\begin{table}[h]

\caption{Proposed designs of a megawatt TE generator ($\dT=10$K) $^2$.  } \label{tab:optmp1}

\vspace{0.2cm}
\centering

\begin{tabular}{|c|c|c|c|c|c|c|c|}
\hline
\hline
 $t_\tube $   & $R_\tube$   & $L_\tube$  &$f_{\max}$   & h& $P_\gain$  & $N_\tube$  \\
\hline
0.1 &  1 & 1&0.43  & 1.11 & 1.39 & 719 \\
\hline

 0.2 &  1 & 1&0.52  & 0.75 & 0.841 & 1189 \\
\hline

 0.5 &  1 & 1&0.63  & 0.43 & 0.408 & 2453 \\
\hline

 1 &  1 & 1 &0.70  &0.28 &0.227 &  4415 \\
\hline
\end{tabular}

\end{table}

\end{center}

\section{Conclusion}
The estimated cost per year of one megawatt electricity in 2016 is about $0.83$ million USD from conventional coal plants to $1.84$ million USD from photovoltic power plants~\cite{DOEoutlook2010}. Also, the investment of the world largest photovoltic farm (Sarnia PV power plant in Canada \cite{FirstsolarSarnia}) is 300 million USD for an annual capacity 1.2$\times 10^5$  MWh, or equivalently, an effective power of 14 megawatt (designed peak power 60 megawatt).  Even if  Table~\ref{tab:optmp1} underestimates  the number $N_\pipe$ of TE tubes by 10 fold, it still appears feasible to manufacture a megawatt TE generator using waters of 10K temperature difference as ``fuel'' which lasts for 20 years and costs less than $20\times 1.84$ million USD, and is hence economically competitive compared with photovoltic farms. For geothermal sources, we anticipate the temperature $\dT$ is at the order of $50$ K or higher, implying an increase of power gain by 25 fold since $P_\gain$ scales as $\dT^2$.

In summary, we have presented a simple design of heat exchanger or electricity generator which use heat resources of small temperature differences. The theoretical analysis shows that it is feasible and economically competitive to build large-scale ocean-based or geothermal power plants by thermoelectric effects.  A for-the-moment science-fiction vision is that power plant units like barges are manufactured on shore and towed to a selected tropical ocean site where electricity is generated and transmitted to shore. Whenever more power is needed, new units can be manufactured, towed and assembled to the site. The renewability and scalability of the TE power plants may yield a way to resolve the energy crisis facing human society.



\appendix

\section{Flow and thermoelectric problems in a TE tube}
\subsection{Flow problem}

 For a low Reynold number  ($Re<2300$)   steady flow in a circular pipe of length $L$ and radius $R$ and maintained by a pressure drop $\dP_\iinn$, the flow is laminar and referred to as the Hagen-Poiseuille flow. Let $v=v(r)$  be the velocity  of the flow, and  by the Navier-Stokes equation we have
\beas
\becs
 \mu \frac{1}{r}\frac{d}{dr} r \frac{d}{dr} v(r)-p'=0 &\iif\; r\in [0, R_\iinn),\\
v(r)=0 &\iif\;r=R_\iinn,
\eecs
\eeas
where $\mu$  is the viscosity of water at $293$K and $p'=-\dP_\iinn/L_\iinn$ is the pressure gradient. The solution to the above problem is given by
\beas
v(r)=\frac{\dP_\iinn (R_\iinn^2-r^2)}{4\mu L_\iinn},
\eeas
and henceforth volumetric flow rate is given by
\beqs \label{eq:Qv}
Q_\iinn=\int_0^{R_\iinn} v(r) 2\pi r dr=\frac{\pi \dP_\iinn  R_\iinn^4}{8\mu L_\iinn}.
\eeqs
The  power dissipated by the viscosity is given by
\beqs \label{eq:aPd}
P^d=   \int_0^{L_\iinn} \int_0^{R_\iinn} \mu |\nabla v|^2 2\pi r dr dx=\frac{\pi\dP_\iinn^2 R_\iinn^4}{8\mu L_\iinn} \nonumber \\
={\dP Q_\iinn}=\frac{8 \mu L_\iinn Q_\iinn^2}{\pi R_\iinn^4}=:\mu \gamma L Q^2/A^2,
\eeqs
where $A$ is the cross-sectional area and $\gamma=8\pi$ for circular pipe. 

\subsection{Thermoelectric problem }

\begin{figure}[t]
\begin{center}
    {\scalebox{1}{\includegraphics{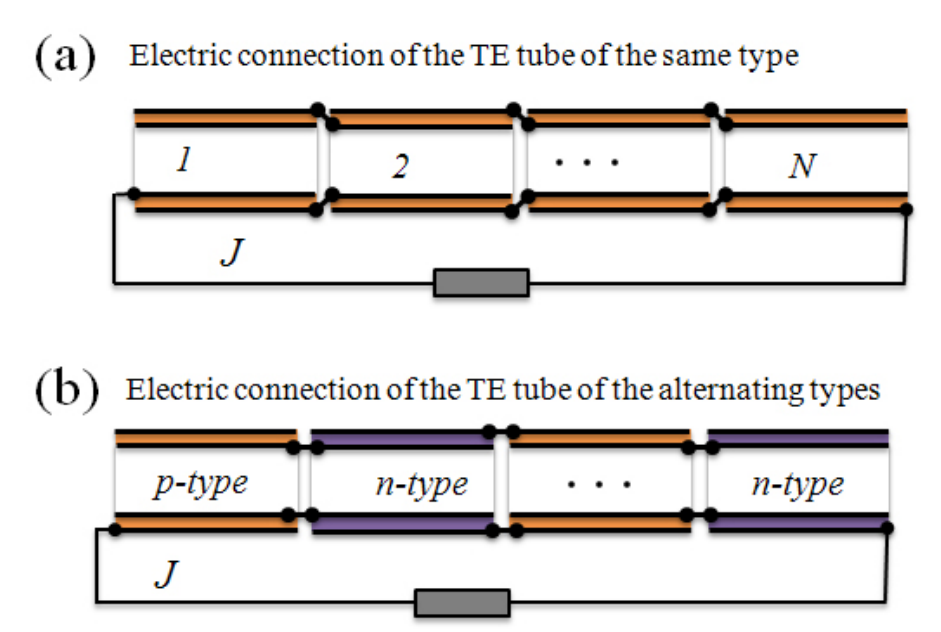}}}

\caption{(a) Electric connection of the TE tube in the longitudinal direction if the type of TE material remains the same; (b) Electric connection of the TE tube in the longitudinal direction for alternating type of if TE materials. }\label{fig:connections}

\end{center}
\end{figure}

 We consider the thermoelectric problem  as water flows through the TE tube immersed in an ambient medium of 10 K temperature difference. The TE tube will be divided into multiple segaments to boost electric potential difference and power output. Assume that each segment of tube is short enough such that the temperature and electric potential on interior and exterior walls may be regarded as constants on each segment, see Fig.~\ref{fig:connections}(a)-(b). Further, it is desirable that each segment of the TE tube works at the optimal conditions to have maximum power generation per unit volume, i.e.,  the segment length $l_i$ and total electric current $J$ shall be controlled  such that the electric current density across the wall of each segment is given by (cf., \S~3.1.1 of Liu, 2012)
\beqs \label{eq:je}
j_e(x)=\frac{ s \sigma (T_\ex-T_\iinn (x_i))}{2t_\tube },
\eeqs
where  the temperature of the exterior walls of all TE tubes is assumed to be constant and equal to $T_\ex$, and $x_i$ is the midpoint of the $i$th segment. Since the  electric current in the closed circuit is constant, the length  of each segment $l_i$ shall be proportional to $1/(T_\ex-T_\iinn(x_i))$.
At this condition, the efficiency of conversion is not optimal and given by
\beqs \label{eq:eta}
\eta=\eta_s \eta_\carnot, \qquad \eta_s=\frac{ZT_0}{4+2ZT_0},
\eeqs
which is only 0.5\% less than the optimal efficiency $\eta_s^{\rm opt}=(\sqrt{ZT_0+1}-1)/(\sqrt{ZT_0+1}+1)$ for $ZT_0 =1$, and even smaller for smaller $ZT_0$.
Also, the potential difference between exterior and interior walls at the $i$th segment is given by
\beqs \label{eq:dmu}
\mu_\ex(x_i)-\mu_\iin(x_i)=-\frac{s}{2}(T_\ex-T_\iinn (x_i)) \quad (i=1,\cdots, N).
\eeqs
The reader is invited to check for the detailed calculations of above results in Liu (2012~\cite{LiuTE2012}).

Moreover, the heat flux flowing out of the tube across the tube wall is given by
\beqs \label{eq:ju}
q(x)=-\kappa_q\frac{T_\ex-T_\iinn(x)}{t_\tube },\quad
\kappa_q=\kappa+T_0\sigma s^2/2.
\eeqs
 For simplicity we assume that the length of the tube is much larger than the cross-sectional diameter of the tube, and hence temperature may be approximately regarded as uniform on each cross-section of the tube. Neglecting the heat generated by viscosity,  by the energy balance  we have
\beqs \label{eq:odeTx}
\rho C_p Q_\tube \frac{d}{dx}T_\iinn(x)+ l_\tube q=0,
\eeqs
where $l_\tube$ is the length of the perimeter of a cross-section of the tube. In account of the inlet temperature $T_\iinn(x=0)={T_\iinn^0}$ which may be assumed to be the temperature of the source, by \eqref{eq:ju} and \eqref{eq:odeTx} we have
\beqs \label{eq:Tinx}
T_\iinn(x)=T_\ex +({T_\iinn^0}-T_\ex ) \exp(-x/L_\ast  ),
\eeqs
where
\beqs \label{eq:Le}
L_\ast  =\frac{\rho C_p  Q_\tube  t_\tube }{ \kappa_q l_\tube}.
\eeqs
Therefore, the outlet temperature is given by
\beqs \label{eq:Tcend}
&&T_\iinn(L_0)=T_\ex +e^{-\lambda/2} ({T_\iinn^0}-T_\ex ),\nonumber \\
&&\lambda=\frac{2L_\tube}{L_\ast  }=\frac{ 2\kappa_q   l_pL_\tube}{\rho C_p  Q_\tube  t_\tube }.
\eeqs
The power generated by each tube is a summation of power generated by all segments and well approximated by (cf., \S~3.1.1 in Liu, 2012)
\beas
&&P_\out=\int_0^{L_\tube} \frac{P_f}{4}\frac{(T_\iinn(x)-T_\ex )^2}{t_\tube ^2} t_\tube  l_\tube dx\\
&&\hspace{0.7cm}=\frac{ P_f \dT^2 l_\tube L_\tube}{4t_\tube   } \cdot
\frac{1-\exp(-\lambda)}{\lambda}.
\eeas

\end{document}